\documentclass[1, 12pt]{elsarticle}
\usepackage[english]{babel}
\usepackage{amsmath}
\usepackage[toc,page]{appendix}
\usepackage{amssymb}        
\usepackage{amsthm}        
\usepackage{natbib}         
\usepackage{mathtools}
\usepackage{graphicx}       
\usepackage{subfigure}      
\usepackage{color}
\usepackage{gensymb} 
\title{Grain boundary strengthening of FCC polycrystals} 

\author{R. A. Rubio$^{1}$}
\author{S. Haouala$^{1}$}
\author{J. LLorca$^{1, 2, }$\corref{cor1}}
\address{$^1$ IMDEA Materials Institute, C/ Eric Kandel 2, 28906, Getafe, Madrid, Spain. \\  $^2$ Department of Materials Science, Polytechnic University of Madrid, E. T. S. de Ingenieros de Caminos. 28040 - Madrid, Spain.}

\cortext[cor1]{Corresponding author}

\begin{document}

\begin{frontmatter}

\begin{abstract}

The effect of grain size on the flow strength of FCC polycrystals was analyzed by means of computational homogenization. The mechanical behavior of each grain was  dictated by a dislocation-based crystal plasticity model in the context of finite strain plasticity and takes into the account the formation of pile-ups at grain boundaries. All the model parameters have a clear physical meaning and were identified for different FCC metals from dislocation dynamics simulations or experiments. It was found that the influence of the grain size on the flow strength of FCC polycrystals was mainly dictated by the similitude coefficient $K$ that establishes the relationship between the dislocation mean free path and the dislocation density in the bulk. Finally, the modelling approach was validated by comparison with experimental results of the effect of grain size on the flow strength of Ni, Al, Cu and Ag.
\end{abstract}


\end{frontmatter}

\section{Introduction}

Plastic deformation of metallic alloys is controlled by dislocation motion and the strategies to increase the flow stress are always based in the introduction of obstacles to dislocation slip. Reducing the grain size in polycrystals is an obvious choice and Hall \cite{H51} and Petch \cite{P53} reported a phenomenological relationship  for different metallic alloys between the yield stress $\sigma_y$, and the average grain size, $\overline{D}_g$, of the form

\begin{equation}
\centering
\sigma _y=\sigma_{\infty} + C_{HP}\overline{D}_g^{-0.5},
\label{hallpetch-initial}
\end{equation}

\noindent where $\sigma_{\infty}$ is the yield stress of a polycrystal with a very large grain size and $C_{HP}$ is a material constant. Further experimental data as well as theoretical investigations confirmed the increase in yield strength when the grain size was reduced although the experimental data were better approximated when the influence of grain size was introduced as $\overline{D}_g^{-x}$ with $ 0.5 \le x \le 1$ \citep{RP86, DB13, DB14, LBD16}.

From the physical viewpoint, the presence of grain boundaries influences the dislocation motion and dislocation storage because slip transfer is not possible across most grain boundaries due to the misorientation of the slip planes, leading to the formation of dislocation pile-ups at the grain boundaries \cite{BEZ14, HNV18, BAL19}. The back-up stress induced by the pile-ups increases the critical resolved shear stress necessary to nucleate and move dislocations within the grain. Another implicit outcome of the misorientation between grain boundaries and of the anisotropy of the plastic deformation in each grain within the polycrystal is the generation of geometrically-necessary dislocations near the grain boundaries. 
They  account for the deformation incompatibility between grains with different orientation and these additional dislocations also contribute to the hardening. Different theoretical models have been proposed in the past to account for the hardening induced by both mechanisms \citep{A70, K70, H72, LBD16} but it is obvious that the overall hardening due to the presence of grain boundaries depends on many factors (the FCC metal, the elastic  and plastic anisotropy of the grains, the initial dislocation density, the presence of other obstacles to dislocation motion, etc.) that cannot be captured by simple theoretical models.

Computational homogenization using crystal plasticity to describe the behavior of each crystal is currently the most powerful strategy to simulate mechanical response of polycrystals \citep{SLL18}.
Grain boundaries are assumed to be transparent for slip transfer in standard crystal plasticity models and, thus, the effect of the grain boundary on the slip transmission is not accounted for.  Lower-order strain-gradient crystal plasticity models can account for the effect grain boundaries through the density of geometrically-necessary dislocations generated near the boundaries due to the deformation incompatibility between grains with different orientations \citep {EPB02, CBA05}. Moreover, higher-order strain-gradient crystal plasticity models can also control the dislocation flux across the boundary by means of the higher-order boundary conditions \citep{LN16}. Nevertheless, strain-gradient crystal plasticity models are computationally very expensive and, thus, they are not suitable for large simulations involving many crystals in a polycrystal. 

Phenomenological crystal plasticity models can introduce the effect of grain boundaries through different parameters that take into account the increase in dislocation density near the grain boundary due to the formation of dislocation pile-ups \citep{W2011, MBB15, SZM16}. Within this framework, Haouala {\it et al.} \citep{HSL18} recently developed a physically-based crystal plasticity model which was able to reproduce very accurately the effect of grain size on the flow strength of polycrystalline Cu. The model assumes the critical resolved shear stress is a function of the dislocation density through the Taylor model \citep{T34}, while the dislocation density evolution during deformation follows a Kocks-Mecking law \citep{KM03}. The effect of grain boundaries is taken into account by including the effect of the distance to the grain boundary in the term of the Kocks-Mecking law that dictates the generation of new dislocations \cite{L06}. In this way, the effect of dislocation pile-ups at the grain boundaries is considered. 

One important feature of this model is that all parameters in the model have a clear physical meaning and could be obtained from either dislocation dynamics simulations in FCC polycrystals and from experimental observations in Cu. In this paper, this simulation strategy is extended to other FCC metals: Ag, Al and Ni. As in the previous paper, the model parameters were obtained from dislocation dynamics simulations or from experimental observations in each metal. The simulations results were validated against experimental data of the effect of grain size on the strength of these FCC metals and, thus, the effect of different material parameters (similitude coefficient, annihilation distance between dislocations, etc.) on grain boundary strengthening was ascertained. This information can be very useful to design new metallic alloys with enhanced grain boundary strengthening.

The outline of the paper is the following. After the introduction, the crystal plasticity model  and the computational homogenization strategy are briefly recalled in sections \ref{sec2} and \ref{sec3}, respectively.  The simulation results and the corresponding comparison with experimental data for Al, Ag and Ni are included in Section \ref{sec4}, while the main conclusions of the paper are summarized in the last section.

\section{Physically-based crystal plasticity model}\label{sec2}

The physically-based crystal plasticity model recently developed by Haouala {\it et al.} \cite{HSL18} has been used as constitutive equation for the grains in this study. The mechanical behavior of each grain is dictated by a rate-dependent crystal plasticity model in the context of finite strain plasticity and takes into the account the formation of pile-ups at grain boundaries. The model description and implementation is described in detail in \cite{HSL18} but the main features are briefly recalled here for the sake of brevity. 

The plastic strain rate in each slip system $\alpha$, $\dot{\gamma}^\alpha$, can be obtained as a function of the corresponding resolved shear stress on the slip plane and slip direction, $\tau^\alpha$, according to \cite{K75, KL78},

\begin{equation}
    \dot{\gamma}^\alpha=\dot{\gamma}_0\left(\frac{| \tau^\alpha |}{\tau^\alpha_c}\right)^{\frac{1}{m}}sgn(\tau^\alpha),
\end{equation}

\noindent where  $m$ stands for the strain rate sensitivity coefficient, $\dot{\gamma}_0$ is the reference shear strain rate and $\tau^\alpha_c$ the critical resolved shear stress  on the slip system $\alpha$. 

It is assumed that the initial value of the critical resolver shear stress is negligible (a reasonable assumption for pure FCC metals) and that increases with the plastic strain due to the interactions between dislocations. Thus,  $\tau^\alpha_c$ is expressed by the generalized Taylor model \citep{FBZ80} according to,

\begin{equation}
    \tau^\alpha_c=\mu b \sqrt{\sum_\beta a^{\alpha \beta}\rho^\beta}
    \label{CRSS}
\end{equation}

\noindent where $\mu$ and ${b}$ stand, respectively, for the shear modulus (parallel to the slip plane) and the Burgers vector of the material,  while $\rho^{\beta}$ accounts for the dislocation density in the slip system $\beta$. $a^{\alpha \beta}$ are the dimensionless coefficients that describe the strength of the different interactions between pairs of slip systems. In the case of FCC materials, they were obtained  from discrete dislocation dynamics simulations \cite{DHK08, BCB13} and  are given in table \ref{table:1}.

\begin{table}
\caption{Parameters of the dislocation-based crystal plasticity model for FCC metals.}
\begin{tabular}{l  l  } \hline
\\
 {\it Viscoplastic parameters}\cite{ZKZ03} & \\
 Reference shear strain rate  $\dot{\gamma}_0$  (s$^{-1}$) & 0.001\\
 Strain rate sensitivity coefficient $m$ & 0.05 \\ \\

 {\it Dislocation interaction coefficients:}\cite{DHK08, BCB13} & \\
Self interaction  & 0.122\\
 Coplanar interaction & 0.122\\
 Collinear interaction & 0.657\\
 Hirth lock & 0.084\\
 Glissile junction & 0.137\\
 Lomer-Cottrell lock & 0.118\\ \\
 
 {\it Grain boundary dislocation storage coefficient} $K_s$   & 5 \cite{SDK10}\\
 \hline
\end{tabular}
\label{table:1}
\end{table}

New dislocations are necessary to maintain the plastic flow during deformation. They are generated as  result of the interaction of dislocations with obstacles. In the case of pure metals, the only obstacles (besides grain boundaries) are other dislocations and the rate at which dislocations are generated depends on the average distance between dislocations. Moreover, dislocations in different slip planes with opposite Burgers vectors are attracted to the each other and can mutually annihilate. This is difficult in the case of edge dislocations because it requires dislocation climb but it is easier in the case of screw dislocations that can cross slip. These mechanisms can be taken into account through the Kocks-Mecking phenomenological model \cite{KM81, KM03}, in which the net storage rate of dislocations per slip system, $\dot{\rho}^\alpha$, is given by 

\begin{equation}
    \dot{\rho}^\alpha=\frac{1}{b}\left(\frac{1}{\emph{l}^\alpha}-2y_c \rho^\alpha \right) | \dot{\gamma}^\alpha |
    \label{KM}
\end{equation}

\noindent where  $\emph{l}^\alpha=\frac{K}{\sqrt{\sum_{\beta\neq\alpha} {\rho^\beta}}}$ is the dislocation mean  free path (MFP) in the system $\alpha$, $\rho^\beta$ are the dislocation densities in the other slip systems, and $K$ is a dimensionless constant (known as the similitude coefficient \citep{SK11}). The negative term  in eq. \eqref{KM} accounts for dynamic recovery and  is proportional to the effective annihilation distance between dislocations, $y_c$. 

The effect of dislocation pile-ups at the grain boundaries was introduced by Haouala {\it et al.} \cite{HSL18} by including another term in eq. \eqref{KM}, which was expressed as 
\begin{equation}
   \dot{\rho}^\alpha=\frac{1}{b}\left(max \left(\frac{1}{\emph{l}^\alpha},\frac{K_s}{d_b}\right)-2y_c\rho^\alpha\right)| \dot{\gamma}^\alpha |,
   \label{KMd}
\end{equation}

\noindent where $d_b$ stands for the distance to the grain boundary and $K_s$ is a dimensionless constant that controls the dislocation storage at the grain boundary due to the formation of pile-ups \cite{L06}.   
$K_s \approx 5$ was determined from 3D dislocation dynamics simulations in the presence of impenetrable grain boundaries in FCC polycrystals \cite{SDK10}. It is assumed that the formation of pile-ups under these circumstances is mainly controlled by the elastic interactions between dislocations and, then, it should depend weakly on temperature and strain rate.

\section{Computational homogenization framework}\label{sec3}

\begin{figure}[h]
    \centering
    \includegraphics [scale=0.4, angle=-90]{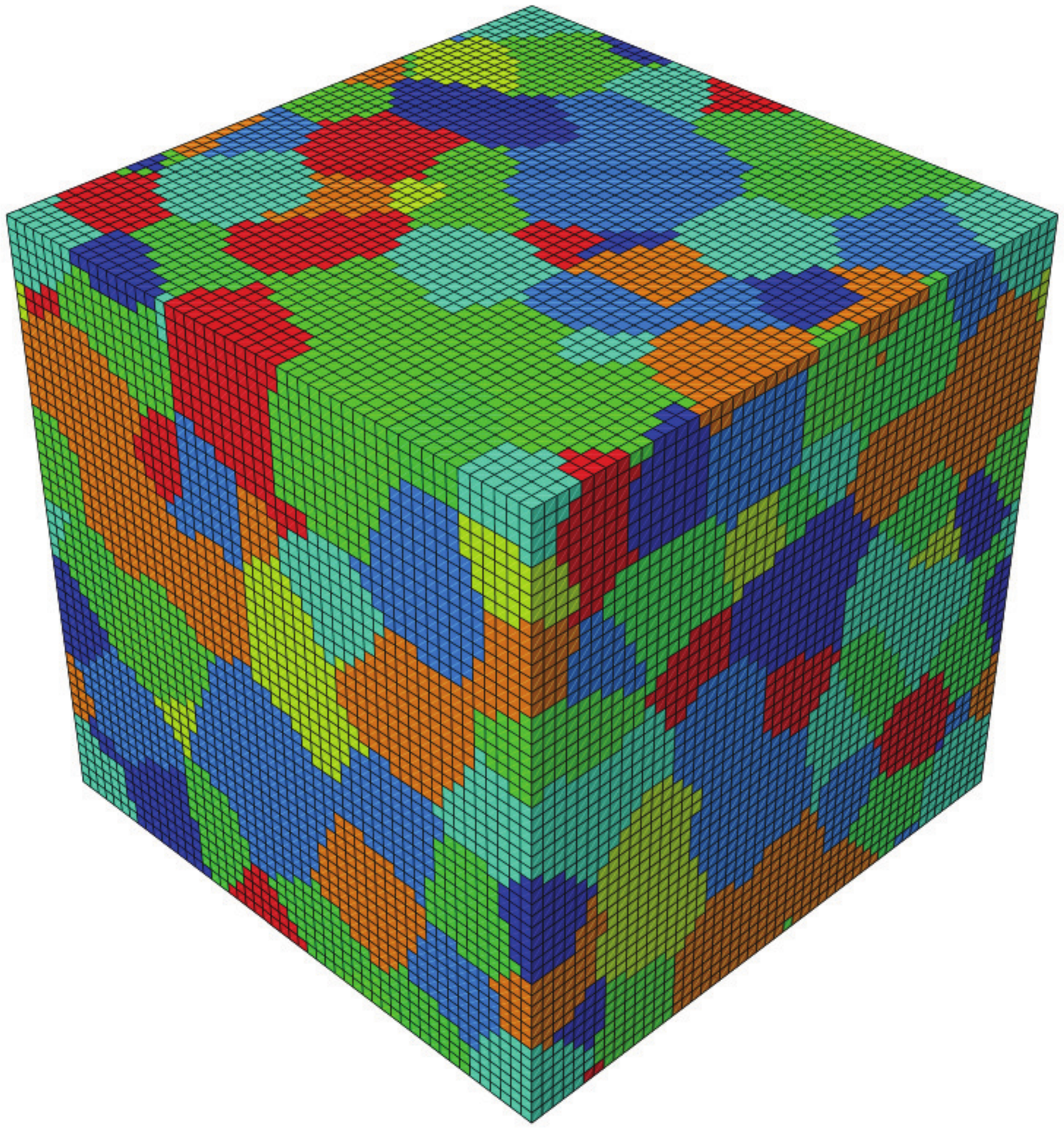}
    \caption{Representative volume element of FCC polycrystal containing 200 crystals discretized with 125000 cubic finite elements.}
    \label{RVE-grains}
\end{figure}

The mechanical behavior in uniaxial tension of polycrystals with different grain size was computed using a standard computational homogenization framework using the finite element method \cite{SLL18}. The microstructure of the polycrystal is represented using a cubic representative volume element with equiaxed grains and random texture, generated using Dream3D  \cite{DREAM3D} (fig. \ref{RVE-grains}). 
The size of each grain was given by the diameter $D$ of a sphere with the same volume and the grain size in the polycrystal followed a log-normal distribution with an average grain size $\overline{D}_g$ and a standard deviation  of 0.2$\overline{D}_g$. 

The representative volume element of the microstructure contained 200 grains and was discretised using $50 \times 50 \times 50$ cubic elements (C3D8 Abaqus elements) (Fig. \ref{RVE-grains}). The microstructure was periodic and periodic boundary conditions were applied to opposite surfaces of the RVE to determine the mechanical response. At the beginning of the simulation, the distance of each Gauss point to the nearest interface along each slip system, $d_b$, was obtained to be used in eq. \eqref{KMd}. These values were assumed to be constant through the deformation because the maximum applied strain was small (5\%).

Simulations were carried out with Abaqus/Standard \cite{A18} within the framework
of the finite deformations theory with the initial unstressed state as reference. More details about the computational homogenization strategy can be found in \cite{HSL18}.

\section{Results and discussion}\label{sec4}

\subsection{Parameter identification}

The crystal plasticity model presented in section \ref{sec2} depends on a number of physical parameters characteristic of each FCC metal. They are identified below in the cases of Al, Ni and Ag, while the corresponding values for Cu can be found in \cite{HSL18}. The first set of values are the elastic constants and the Burgers vector, that can be found in the literature \cite{Al-elastic-constants, Ni-elastic-constants, Ag-elastic-constants}. They are given in Table \ref{table:2}, together with the shear modulus $\mu$ parallel to the slip plane, which can be computed from the elastic constants.

\begin{table}[h]
\caption{Parameters for Al, Ni and Ag single crystals of the dislocation-based crystal plasticity model}
\begin{tabular}{ l  l  l  l  l  l  }
 
 \hline
 & Al & Ni & Ag\\
  \hline
 \it{Elastic constants (GPa)} &&&\\
 $C_{11} $ & 108   & 249  & 124 \\
 $C_{12} $ & 61.3  & 155  & 93.7 \\
 $C_{44} $ & 28.5  & 114  & 46.1 \\
 $\mu$  & 25.0  & 58.4   & 19.5 \\ \\

 \it{Dislocation parameters} & & & \\
 Burgers vector $b$ (nm): & 0.286  & 0.250  & 0.288 \\
 Annihiliation distance edge dislocations $y_e$ (nm)  & 6$b$ \cite{MWW00} & 6$b$  \cite{MWW00}& 6$b$ \cite{MWW00} \\
 Annihilation distance  screw dislocations  $y_s$ (nm)   & $113$ \cite{AP02} & $26.5$ \cite{PKB13} & $23.5$ \cite{PKB13}\\
 Effective annihilation distance  $y_c$ (nm)  & 56 & 14 & 12.5 \\
Similitude coefficient $K$ & 9 \cite{SK11} & 11 \cite{SK11} &5 \cite{SK11}\\
 \hline
\end{tabular}
\label{table:2}
\end{table}

The similitude coefficient $K$ is a dimensionless constant that arises from the  experimental observation (known as the  similitude principle) that relates the flow stress $\tau$ with the average wavelength of the characteristic dislocation pattern $d$ according to $\tau = K \mu b/d$. This experimental relationship holds for monotonic and cyclic deformation of FCC metals  and  the similitude coefficients under both conditions, $K$ and $K_c^*$, were carefully analyzed in \cite{SK11}. The actual values of the similitude coefficient $K$ under monotonic tension found in the literature are in the range 5 $\le K \le 10$  and the large scatter is associated with the experimental difficulties to measure the dislocations densities and to determine the average diameter of the dislocation-free regions \cite{RP86}. However, the similitude coefficient under cyclic deformation (that relates the saturation stress with the channel width between persistent slip bands) could be measured more accurately for Ni and Al  and Sauzay and Kubin \cite{SK11} reported that  2 $\le K/K_c^* \le$ 2.76. From the $K_c^*$ values reported in \cite{SK11}, $K$ was determined for Ni and Al assuming $ K= 2.76 K_c^*$ and the corresponding values are found in Table \ref{table:2}. There were not enough experimental data to provide a reliable value of $K_c^*$ for Ag but it seemed to be  smaller than those reported for Ni and Al and it was assumed that $K$ = 5, which is in the lower end of the experimental data reported in \cite{RP86}.

The last parameter of the crystal plasticity model is the annihilation distance between dislocations, $y_c$, which depends on the dislocation character. In the case of FCC materials, the annihilation distance between edge dislocations, $y_e$, can be approximated by $6b$ \cite{MWW00}. In the case of screw dislocations, the annihilation distance, $y_s$, depends on the ability of  screw dislocations to cross-slip, a thermally-activated phenomenon which depends on the stacking fault energy of the crystal and also on temperature, strain rate and applied stress. 
The values of $y_s$ for Al single crystals at ambient temperature were taken from the experimental observations in  \citep{AP02}. Regarding Ni and Ag, they were estimated from the  dislocation dynamics simulations in \citep{PKB13}  to predict the saturation stress under cyclic deformation. 
They are summarized in Table \ref{table:2}. Assuming equivalent fractions of edge and screw dislocations,  an effective annihilation distance $y_c = (y_e + y_s)/2$ was used for each FCC crystal in the crystal plasticity model.

\subsection{Effect of grain size on the flow strength}

The effect of the grain size on the tensile deformation of Al, Ag and Ni polycrystals was analyzed using the computational homogenization framework and the crystal plasticity models presented above. Four different simulations were carried out for each material, with average grain sizes $\overline{D}_g$ of 10, 20, 40 and 80 $\mu$m. The initial dislocation density in each slip system in all simulations was 10$^{11}$ m$^{-2}$, leading to a total initial dislocation density $\rho_i=1.2\times10^{12}$ m$^{-2}$.  All the simulations were performed under quasi-static strain rates ($\approx$ 10$^{-3}$ s$^{-1}$).

\begin{figure}[h!]
    \centering
    \includegraphics[scale=0.8]{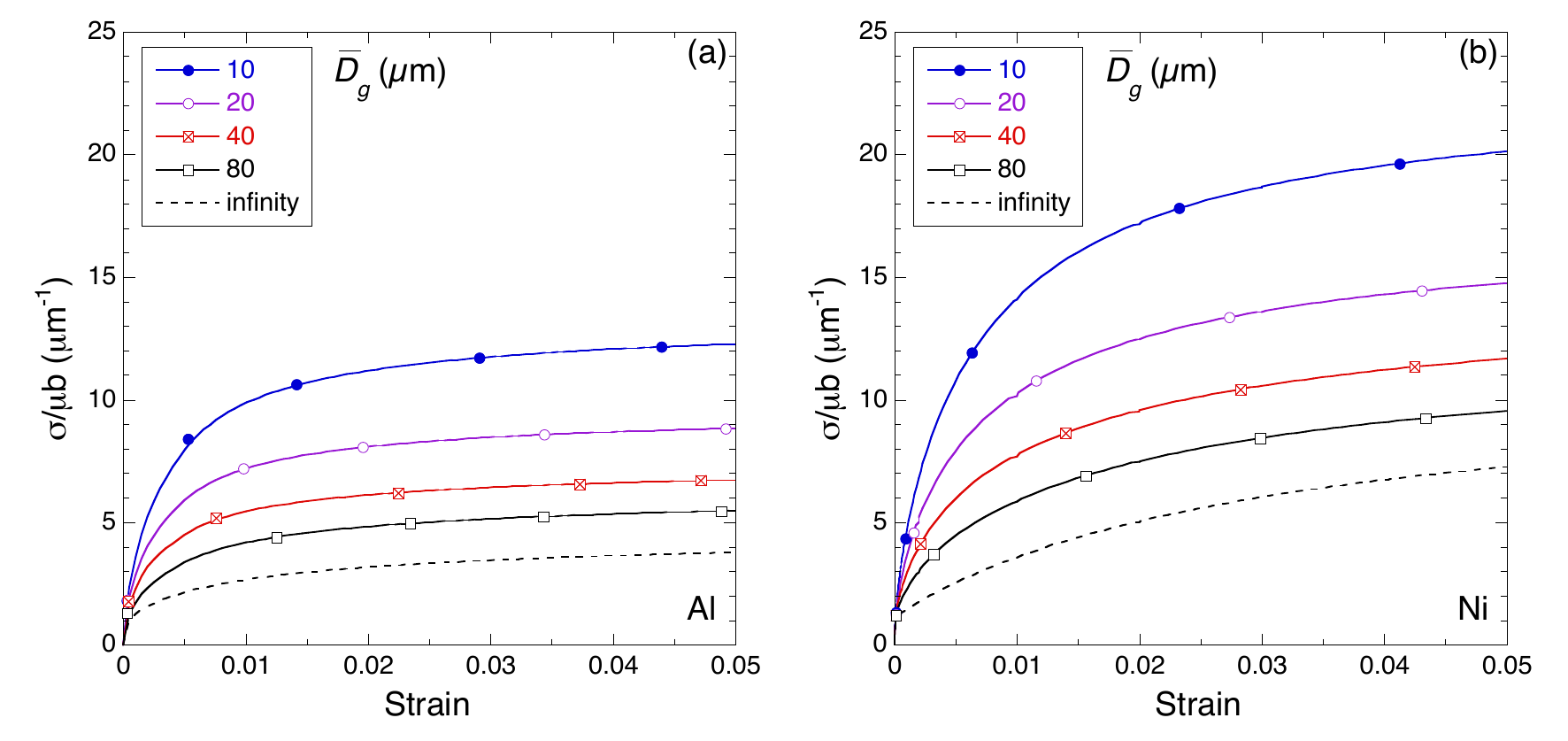}
        \includegraphics[scale=0.8]{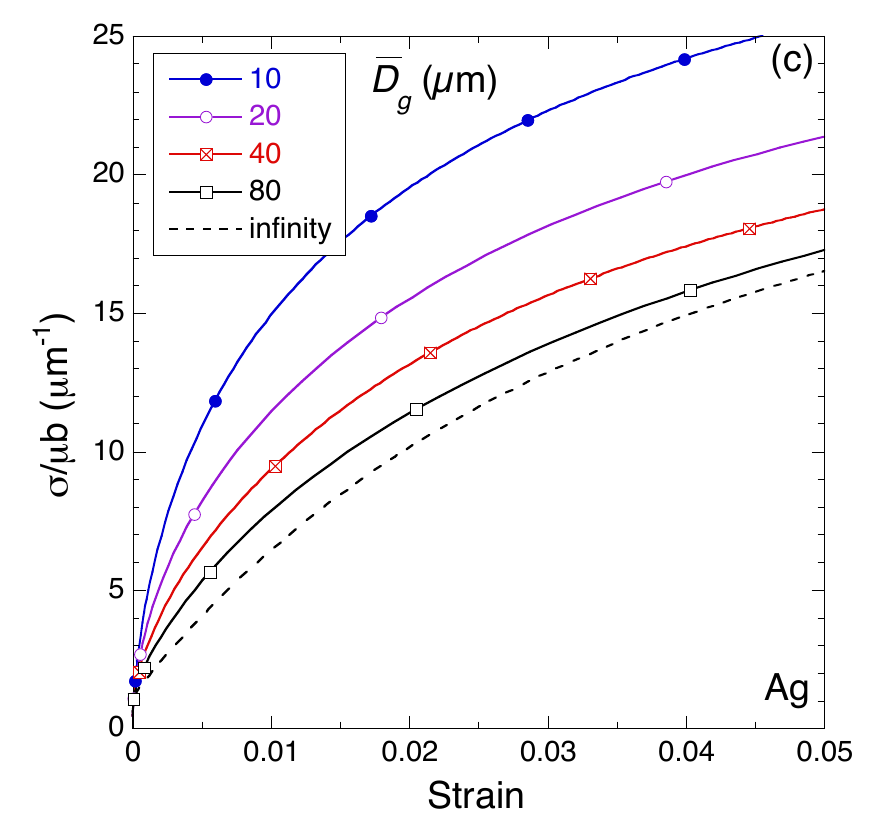}
        \caption{Stress-strain curves of different FCC polycrystals as a function of the average grain size, $\overline{D}_g$.  (a) Al, (b) Ni and (c) Ag. The initial dislocation density was $\rho_i=1.2\times10^{12}$ m$^{-2}$ in all cases. The symbols on the curves are just to distinguish the results for different grain sizes.}    \label{SE}
\end{figure}

\begin{figure}[h!]
    \centering
    \includegraphics[scale=0.8]{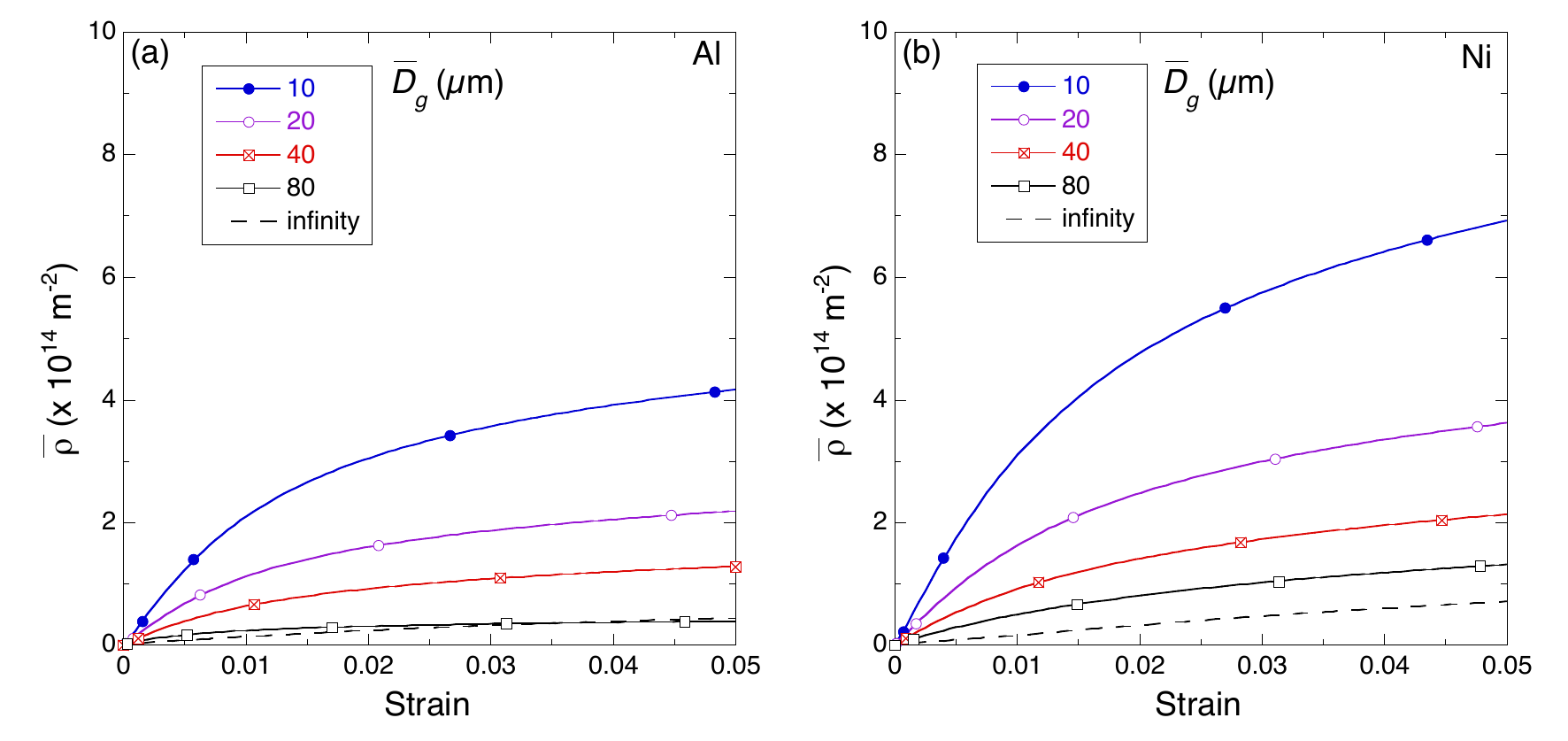}
        \includegraphics[scale=0.8]{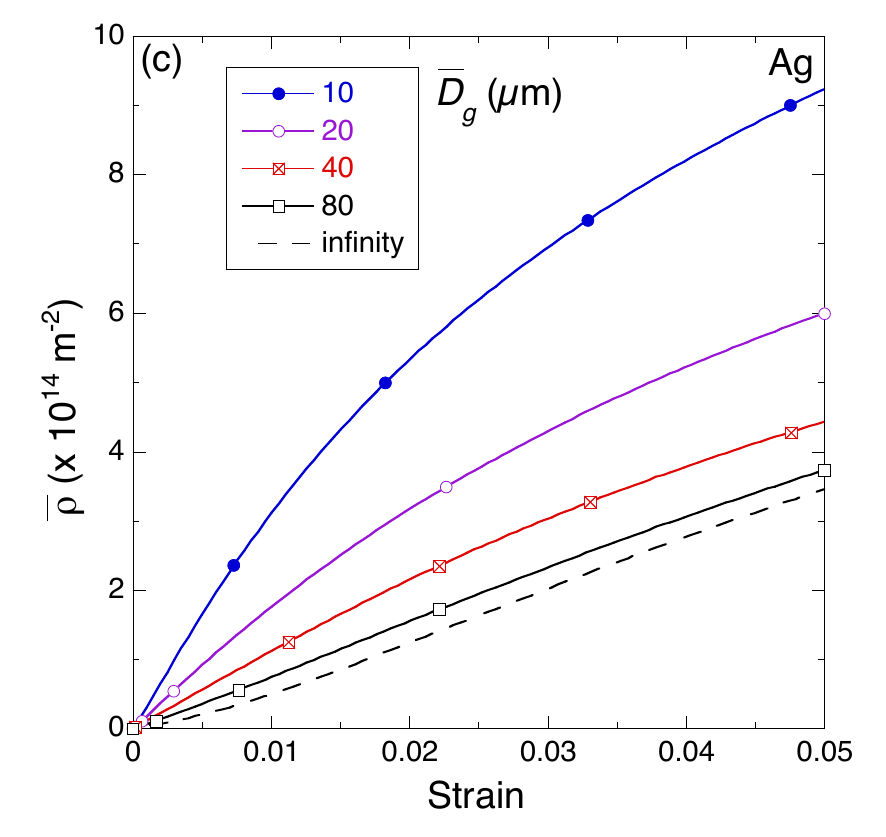}
        \caption{ Evolution of the average dislocation density, $\bar\rho$ of different FCC polycrystals as a function of the average grain size, $\overline{D}_g$.  (a) Al, (b) Ni and (c) Ag. The initial average dislocation density was $\rho_i=1.2\times10^{12}$ m$^{-2}$ in all cases. The symbols on the curves are just to distinguish the results for different grain sizes.}    \label{DDE}
\end{figure}

The engineering stress-strain ($\sigma-\epsilon$) curves  are plotted in Fig. \ref{SE}. An additional simulation was performed for each material using the dislocation evolution law of eq. \eqref{KM}, instead of eq. \eqref{KMd}. These simulations do not take into account the effect of dislocation pile-ups at the grain boundaries and, thus,  stand for the behavior of a polycrystal with "infinite" grain size. The stresses in each plot were normalized by the corresponding values of $\mu b$ to reveal more clearly the differences among the different FCC crystals as a result  of interactions, accumulation and annihilation of dislocations. The stress at the onset of plastic deformation is the same for the three metals  (when normalized by $\mu b$) and different grain sizes and only depends on the initial dislocation density, as dictated by eq. \eqref{CRSS}. Afterwards, the strain hardening depends on three different phenomena: the strain hardening induced by the creation of new dislocations in the bulk and by the dislocation pile-ups at the grain boundaries and the strain softening due to the annihilation of dislocations. The evolution of the average dislocation density, $\bar\rho$, in each RVE with the applied strain is plotted in Fig. \ref{DDE}.

The influence of the dislocation multiplication and annihilation in the bulk can be analyzed for each metal  in the stress-strain and  average dislocation density-strain curves for "infinite" grain size in Figs. \ref{SE}  and Figs. \ref{DDE} (discontinuous lines).  Dislocation multiplication is responsible for the initial strain hardening, which increased as the similitude coefficient $K$ decreases (because the dislocation MFP is proportional to $K$). Thus, the initial strain hardening of Ag was much higher than that of Ni and Al in the crystals with infinite grain size because of the faster dislocation accummulation rate. The average dislocation density increased with the applied strain and dislocation annihilation also influenced markedly the strain hardening rate at higher strains. The annihilation distance for screw dislocations was very large in Al (owing to its high stacking fault energy) (Table \ref{table:2}) and the strain hardening rate has decreased almost to 0 at low strains ($\epsilon \approx 2\%$) while the average dislocation density remained very low (Fig. \ref{DDE}a). This behavior indicates that a hardening-recovery equilibrium was almost reached and the rate of generation of mobile dislocations to maintain the plastic flow was balanced by the rate of dislocation annihilation. On the contrary, the effective annihilation distances of Ni and Ag were smaller than that of Al, and the  steady-state situation was not reached at $\epsilon \approx 5\%$.  In particular, the strain hardening rate  in Ag was still very large at this strain because the generation of new dislocations was still dominant with respect to the dislocation annihilation mechanism due to the small value of dislocation MFP (Fig. \ref{DDE}c).

The presence of grain boundaries increased significantly the flow stress of all the FCC polycrystals, leading to the Hall-Petch behavior (Fig. \ref{SE}). The initial strain hardening rate increased as the grain size decreased and large differences in the flow stress as a function of the grain size were found for $\epsilon$ = 0.5\% as a result of the formation of dislocation pile-ups at the grain boundaries, which increased the average dislocation density (Fig. \ref{DDE}). Nevertheless, dislocation annihilation became noticeable as the dislocation density at grain boundaries increased, leading to a rapid reduction in the accumulation of dislocations and strain hardening rates. The influence of dislocation annihilation increased with the applied strain and was also more noticeable in polycrystals with smaller grain size. As a result, the dislocation accumulation rates and the stress hardening rates at an applied strain of $\approx $ 5\% were practically independent of the grain size and only dependent on the metal. They were  minimum in Al (due to the large values of $K$ and $y_c$) and maximum in Ag (due to the small magnitude of  $K$ and $y_c$).

The strain hardening due dislocation pile-up at the grain boundaries depended on the capacity of the material to  generate and store dislocations at the grain boundary as compared with the capacity to generate dislocations in the bulk. The former is controlled by the dislocations storage coefficient, $K_s$, which was assumed to be the same for all FCC metals. The latter depends on the similitude coefficient $K$ that was high in Ni and Al and low in Ag. Thus, the dislocation MFP and the increase in dislocation density with strain in the bulk was low in Ni and Al and large in Ag. As a result, the initial strengthening due to the formation of grain boundary pile-ups was more noticeable in Ni and Al, as compared with Ag.

The softening induced by dislocation annihilation tended to reduce the high dislocation densities near the grain boundaries and was controlled by the effective annihilation distance $y_c$. It was maximum in the case of Al and its influence is clearly observed in the stress-strain curves in Fig. \ref{SE}a) because the strain hardening rate for $\epsilon > $ 2\% was very low and  equivalent to the one in the polycrystal with "infinite" grain size. Thus, the hardening contribution due to the formation of pile-ups at the grain boundaries is compensated by the dislocation annihilation when $\epsilon > $ 2\% and the Al polycrystals have reached a hardening-recovery equilibrium in which the effect of the grain size on the strain hardening disappears. In fact, the dislocation accumulation rate in Al is practically independent of the average grain size for $\epsilon > $ 2\%, as shown in Fig. \ref{DDE}a). The effective annihilation distance in Ni and Ag was approximately one half of the one in Al and the strain hardening rate of the polycrystals with "infinite" grain size was equivalent to that of the polycrystals with different grain sizes when $\epsilon = $ 5\%. In the case of Ni, the strain hardening rate was almost zero (indicating that hardening-recovery equilibrium was almost reached)  and this conclusion is supported by the plot of the evolution of the average dislocation density with the applied strain in Fig. \ref{DDE}b), which shows that accumulation rate of dislocations was very low for all grain sizes when $\epsilon = $ 5\%. On the contrary,  the hardening rate in Ag was still noticeable at this strain but independent of the grain size.  These results indicate that strain hardening in Ag at $\epsilon = $ 5\% was controlled by the generation of dislocations in the bulk and not by the storage of more dislocations at the grain boundaries.

\begin{figure}[!]
    \centering
    \includegraphics[scale=0.3, angle=0]{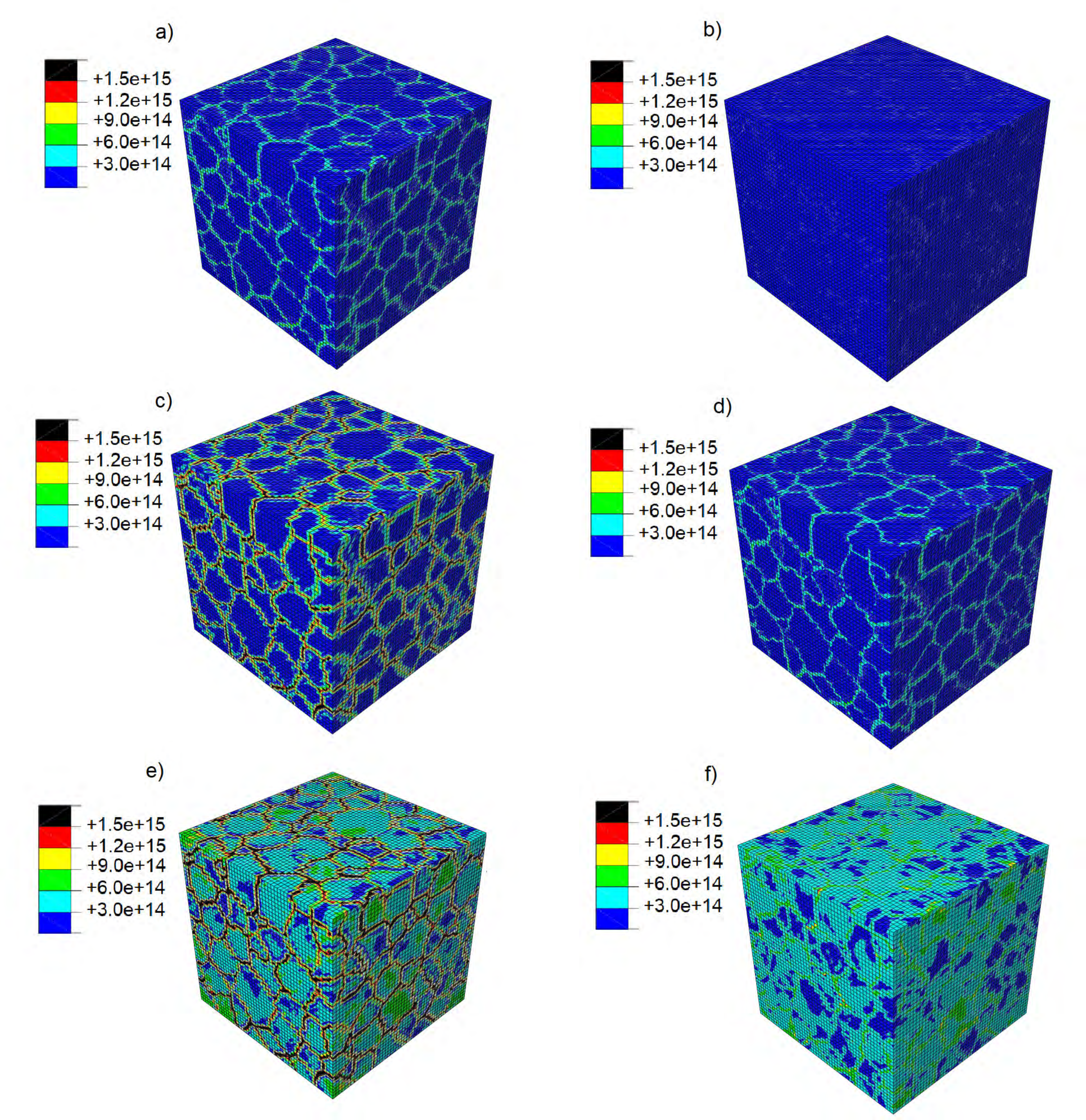}
    \caption{Contour plots of the total dislocation density (expressed in m$^{-2}$) in polycrystals with different average grain size, $\overline{D}_g$  deformed in tension up to 5\% strain. (a) Al, $\overline{D}_g$ = 10 $\mu$m. (b) Al, $\overline{D}_g$ = 40 $\mu$m. (c) Ni, $\overline{D}_g$ = 10 $\mu$m. (d) Ni, $\overline{D}_g$ = 40 $\mu$m. (e) Ag, $\overline{D}_g$=10 $\mu$m. (f) Ag, $\overline{D}_g$ = 40 $\mu$m. The initial dislocation density was $\rho_i=1.2\times10^{12}$ m$^{-2}$ in all cases.}
    \label{fig:DD}
\end{figure}

These mechanisms are illustrated by the contour plots of the total dislocation density shown in Fig. \ref{fig:DD} for Al, Ag and Ni polycrystals with average grains sizes of 40 $\mu$m and 10 $\mu$m deformed up to 5\%. The enhanced dislocation density at the grain boundaries is immediately observed in each plot as well as the effect of average grain size: the smaller the grain, the larger the maximum dislocation density near the grain boundaries. For a given grain size, the smallest dislocation densities in the bulk and in the grain boundaries are always found in the Al polycrystals, as a result of the large effective annihilation distance (Figs. \ref{fig:DD}a and b). The comparison of the contour plots of Ni and Ag (which have similar values of the effective annihilation distance)  show that -- for a given grain size -- the dislocation density in the interior of the Ag grains is higher than that in the interior of the Ni grains because the similitude coefficient  $K$ (and, thus, the dislocation MFP) of Ag is smaller than that of Ni. The maximum dislocation densities near the grain boundaries are, however, very similar for both Ni and Ag for a given grain size because their effective annihilation distances for dislocations were also similar.

\begin{figure}[h!]
    \centering
    \includegraphics[scale=0.22]{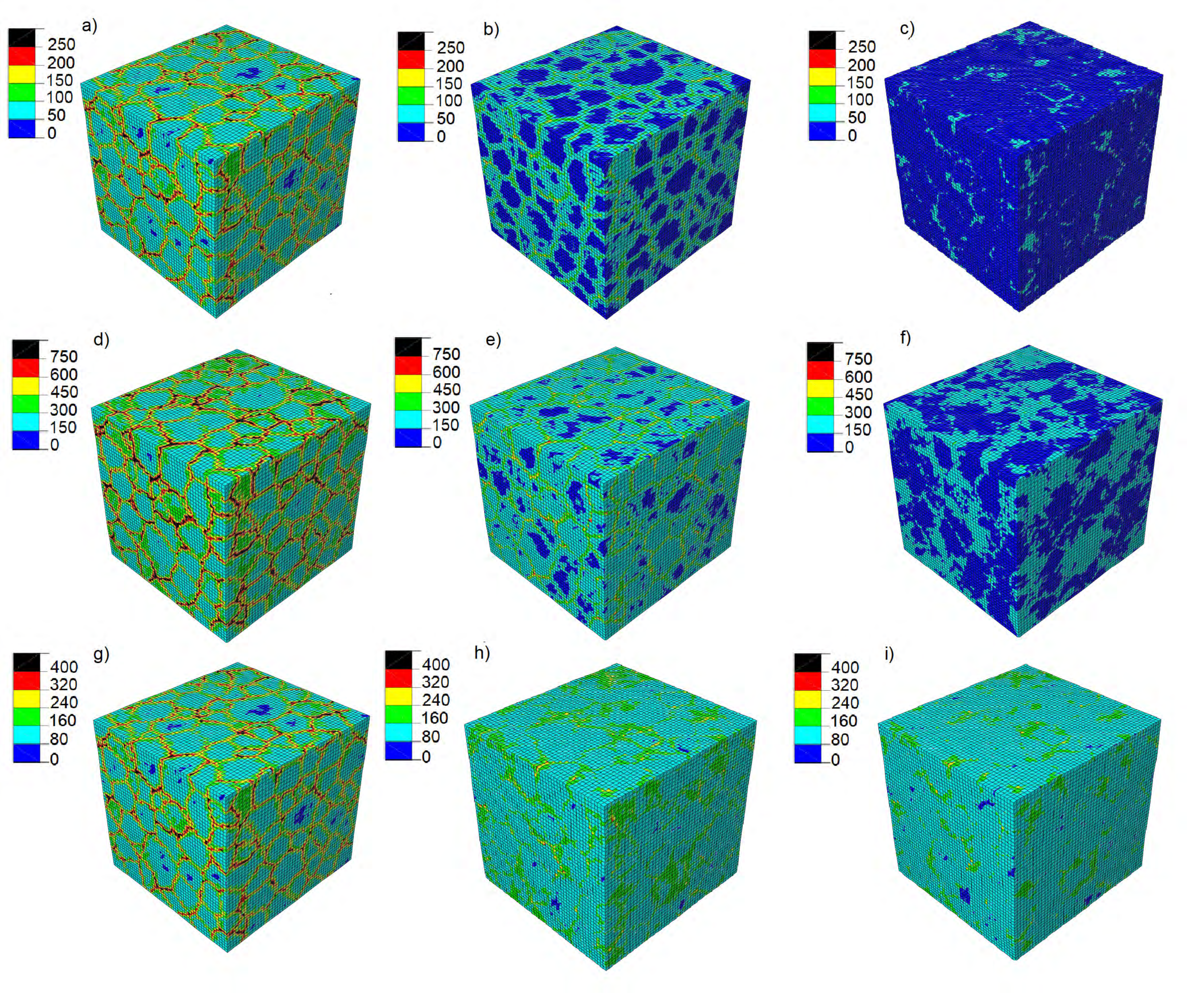}
    \caption{Contour plots of the Von Mises stress in polycrystals with different average grain size, $\overline{D}_g$  deformed in tension up to 5\% strain. (a) Al, $\overline{D}_g$ = 10 $\mu$m. (b) Al, $\overline{D}_g$ = 40 $\mu$m. (c) Al, "infinite" grain size. (d) Ni, $\overline{D}_g$ = 10 $\mu$m. (e) Ni, $\overline{D}_g$ = 40 $\mu$m. (f) Ni, "infinite" grain size. (g) Ag, $\overline{D}_g$=10 $\mu$m. (h) Ag, $\overline{D}_g$ = 40 $\mu$m. (i) Ag, "infinite" grain size. The initial dislocation density was $\rho_i=1.2\times10^{12}$ m$^{-2}$ in all cases. Stresses are expressed in MPa}
    \label{fig:VM}
\end{figure}

Following the Taylor model, eq. \eqref{CRSS}, the maximum stresses in the microstructure are found in the regions with the highest dislocation densities. The contour plots of the Von Mises stress in the Al, Ni and Ag polycrystals deformed up to 5\% are shown in Fig. \ref{fig:VM} for three different average grain sizes: 10 $\mu$m, 40 $\mu$m and "infinity". The contour plots of the polycrystals with "infinite" grain size in Figs. \ref{fig:VM}c, f and i show fairly homogeneous stress fields. The small stress concentrations at the grain boundaries are due to the incompatibility during the elasto-plastic deformation of grains with different orientation. Nevertheless, the formation of dislocation pile-ups at the grain boundaries led to large stress concentrations (that are clearly visible in the contour plots of the Von Mises stress in Fig. \ref{fig:VM}), which increased as the grain size decreased, leading to the Hall-Petch effect. Although the contour plots are similar for the different FCC metals with equivalent grain size, it should be noticed that the differences in stress between the grain boundaries and the bulk were higher in Ni and Al as compared to Ag.

\begin{figure}[h!]
    \centering
    \includegraphics [scale=0.8]{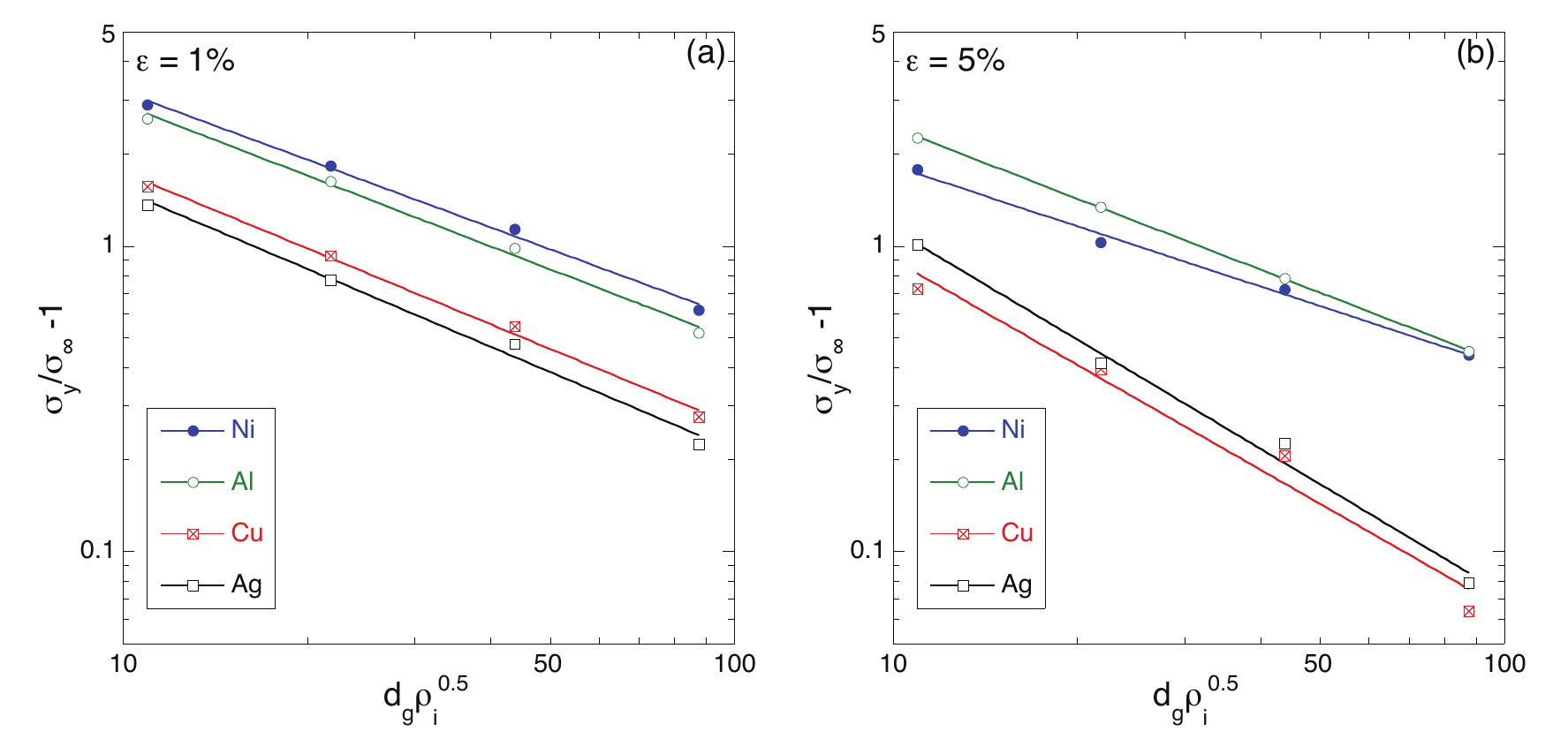}
    \caption{Grain boundary strengthening in FCC metals (Ni, Ag, Al and Cu) as  a function of the dimensionless parameter $\overline{D}_g\sqrt{\rho_i}$. (a) $\epsilon$ = 1\%. (b) $\epsilon$ = 5\%. }
     \label{fig:GB}
\end{figure}

The results presented above (Fig. \ref{SE}) indicate the strengthening provided by grain boundaries depends on the parameters that dictate the generation and annihilation of dislocations of the FCC metal (similitude coefficient $K$, effective annihilation distance $y_c$). Based on the theoretical results \citep{ZS14} and dislocation dynamics simulations \cite{E15}, it was recently proposed that the strengthening due to grain boundaries, $\sigma_y-\sigma_\infty$,  scales with the average grain size, $\overline{D}_g$ and the initial dislocation density in the polycrystal,  $\sqrt{\rho_i}$, according to \citep{HSL18},

\begin{equation}
\sigma_y/\sigma_\infty -1 =  C (\overline{D}_g\sqrt{\rho_i})^{-x}\label{eq:SE}
\end{equation}

\noindent where $\sigma_\infty$ is the flow stress for a polycrystal with "infinite" grain size and $C$ and $x$ are material constants. They are a function of the physical parameters of the model for each FCC metal, namely the elastic constants, the similitude coefficient, the Burgers vector and the annihilation distances between edge ($y_e$) and screw ($y_s$) dislocations. $y_s$ depends on the ability of screw dislocations to cross-slip and it expected to decrease as the temperature increases. The strengthening induced by the grain boundaries in Al, Ni and Ag has been plotted in Figs. \ref{fig:GB}a and b for applied strains $\epsilon$ = 1\% and 5\%, respectively, together with results reported in \citep{HSL18} for polycrystalline Cu\footnote{The similitude coefficient of Cu was $K$ = 6 and the effective annihilation distance $y_c$ = 15 nm.}. The initial dislocation density in all cases was $\rho_i=1.2\times10^{12}$ m$^{-2}$. The results of the numerical simulations for $\epsilon$ = 1\%  (Fig. \ref{fig:GB}a) indicate that the grain boundary strengthening followed eq. \eqref{eq:SE} and the exponent $x$ increased slightly from 0.74 in Ni to 0.85 in Ag. The grain boundary contribution to the strength increased with the similitude coefficient $K$ and was maximum for Ni and Al and minimum for Cu and Ag. 

The results of the simulations at $\epsilon$ = 5\% (Fig. \ref{fig:GB}b) show the effect of the annihilation of dislocations, which reduced the grain boundary strengthening and increased the magnitude of the exponent $x$ in eq. \eqref{eq:SE}. The smallest differences between the grain boundary strengthening at $\epsilon$ = 1\% and 5\% were found for Al because the hardening-recovery equilibrium was practically attained at 1\% and the strain hardening rate was close to zero for larger strains, regardless of the grain size. On the contrary, the largest reduction in the grain boundary strengthening with the applied strain was found in Cu and Ag.

\subsection{Comparison with experiments}

In order to assess the validity of the simulations presented above, the results of the numerical simulations for the flow stress as a function of grain size were compared with experimental data in the literature for Al \citep{al-exp-dat}, NiÊ\citep{ni-exp-dat} , AgÊ\cite{ag-exp-dat} and Cu \citep{HR82}. Experiments were carried out at room temperature under quasi-static loading conditions and thus the effect of the strain rate on the flow stress could be neglected. Nevertheless, the experimental information about the initial dislocation density was not available and it was assumed that it was equal to $1.2\times10^{12}$ m$^{-2}$, a reasonable assumption for a well-annealed polycrystal.

\begin{figure}[h!]
    \centering
    \includegraphics[scale=1.0]{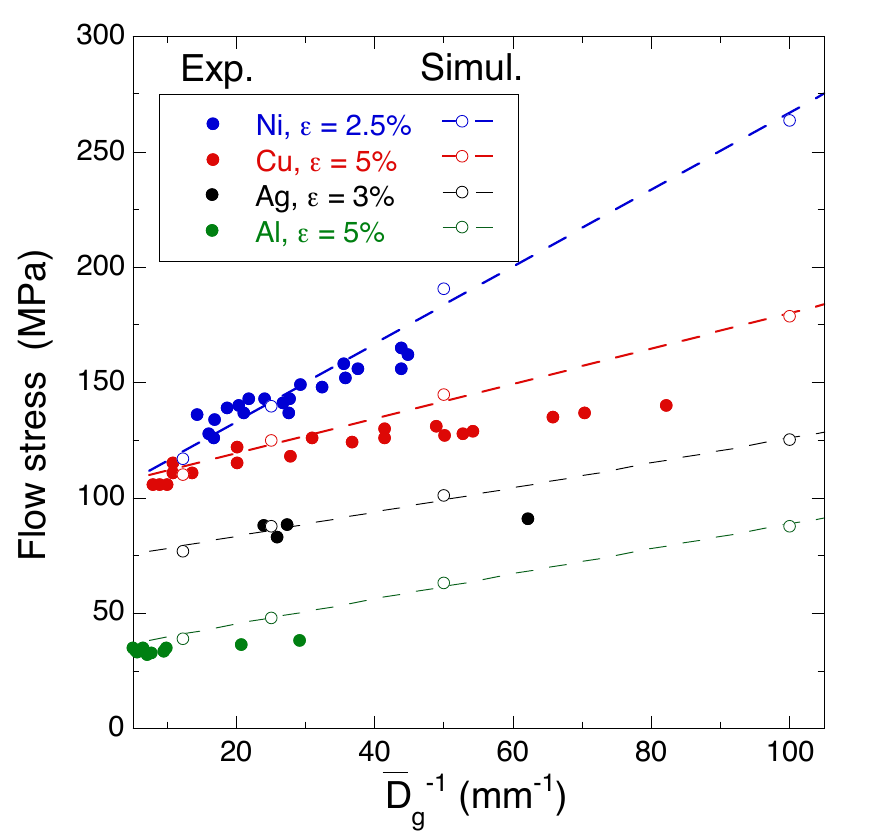}
    \caption{Experimental and simulations results of the flow stress of Al, Ni and Ag polycrystals as a function of the inverse of the average grain size, $\overline{D}^{-1}_{g}$. Solid symbols stand for experimental results and open symbols for simulation results. The flow stresses were measured or computed at different applied strains indicated in the legend.}
    \label{fig:ES}
\end{figure}

The experimental results and the numerical simulations of the flow stress are plotted in Fig. \ref{fig:ES} as a function of the inverse of the average grain size, $\overline{D}^{-1}_{g}$, for Ni, Cu, Ag and Al. The flow stresses were measured or computed at different applied strains indicated in the legend. The agreement was good in all cases indicating that the physically-based crystal plasticity model was able to capture the main strengthening mechanisms due to the accumulation and annihilation of dislocations in the bulk and at the grain boundaries as well as the differences among FCC polycrystals. In general, the numerical simulations tend to overestimate the flow stress for small grain sizes ($\overline{D}^{-1}_{g} > 50$ mm$^{-1}$ or $\overline{D}_{g} <$ 20 $\mu$m) and this behavior may be explained by the fact that the grain boundary strengthening model  in the constitutive equation assumes that all grain boundaries store dislocations and does not take into account  the orientation of the crystals across the grain boundary. Nevertheless,  slip transfer between neighbour grains can be easily accommodated in the case of grains with small misorientation. In addition, the finite element crystal plasticity model may not represent adequately the inhomogeneous plastic deformation that occurs in small grains (below 10 $\mu$m) with low dislocation densities because the voxel size is equivalent to the average dislocation distance.

Regardless of the general agreement of the model with the experimental data, the limitations of the comparison should also be emphasized. The first limitation arises from the experimental data, which do not include information about the initial dislocation densities or the density of annealing twins in the microstructures. Regarding the first point, it was assumed that all the polycrystals were well-annealed and, thus, the initial dislocation density was low in all cases. Annealing twins (often found in Ni, Cu and Ag) are limited by coherent twin boundaries, which are also an obstacle for dislocation transmission. Thus, they can reduce the effective average grain size and lead to higher flow stresses. Twin boundaries could be easily included in the computational homogenization framework as any other grain boundary but they were not because the experimental information was not available. In any case, it should be noted that increasing the initial dislocation density or including coherent twin boundaries in the RVEs would have increased the flow strength of the polycrystals.

Another limitation of the predictive capability of the model comes from the difficulties to obtain accurate estimates of some of the physical parameters that control the mechanical response. They are the similitude coefficient, $K$, the critical annihilation distance for screw dislocations, $y_c$, and the dislocation storage coefficient at the grain boundaries, $K_s$. While the values for these parameters used in the simulations were reasonable and obtained from either dislocation dynamics simulations or experimental data (see section 4.1), there was some uncertainty in the estimates. Moreover, $y_c$ is known to depend on temperature and strain rate because dislocation cross-slip is thermally activated but the only estimates available in the literature are found for quasi-static deformation at ambient temperature. Thus, further experimental and theoretical work is required to provide more accurate results for these parameters as a function of the temperature and strain rate.

\section{Conclusions}

The effect of grain boundaries on the flow strength of FCC polycrystals has been analyzed by means of computational homogenization using a dislocation-based crystal plasticity model. The critical resolved shear stress is a function of the dislocation density through the Taylor approximation, while the generation and annihilation of dislocations follows the Kocks-Mecking model. The effect of dislocation pile-ups at the grain boundaries is included through an additional term in the Kock-Meckings model which is a function of the distance to the grain boundary.  All the model parameters have a clear physical meaning and were identified for each FCC metal from dislocation dynamics simulations or experiments. 

The results of numerical simulations provided a deeper understanding of the influence of the physical parameters that control the plastic deformation of each metal on the Hall-Petch effect. The largest grain boundary strengthening was found in materials with large similitude coefficient $K$ (such as Ni and Al) while the contribution of grain boundaries to the strength was smaller in metals with small similitude coefficient (such as Cu and Ag). The dislocation mean-free path (that controls the generation of dislocations in the bulk) is proportional to $K$ and FCC metals with large similitude coefficients and "infinite" grain size show limited strain hardening. The dislocation accumulation at grain boundaries increased very much the strain hardening rate in these materials, leading to a marked effect of grain boundaries on the flow stress. The effect of grain boundaries on the strain hardening rate was also important in FCC metals with low values of $K$, but these materials also presented a large strain hardening in the absence of dislocation pile-ups and the overall contribution of grain boundaries was lower.

Finally, the model predictions for the effect of grain size on the flow strength were compared with experimental data in the literature for Ni, Cu, Ag and Al. They were in good agreement, particularly for grain sizes $>$ 20 $\mu$m, validating the simulation strategy. 

\section*{Acknowledgements}
This investigation was supported by the European Research Council under the European Union Horizon 2020 research and innovation programme (Advanced Grant VIRMETAL, grant agreement No. 669141).

%\bibliographystyle{elsarticle-num}

%\bibliography{mybibfile}

\end{document}